\documentclass[twocolumn,showpacs,preprintnumbers,amsmath,amssymb]{revtex4}
\usepackage{graphicx}
\def\apj #1 #2 #3 {#1, ApJ, {\bf #2}, #3}
\def\apjl #1 #2 #3 {#1, ApJ, {\bf #2}, L#3}
\def\apjs #1 #2 #3 {#1, ApJS, {\bf #2}, #3}
\def\aap  #1 #2 #3 {#1, A\&A, {\bf #2}, #3}
\def\mnras #1 #2 #3 {#1, MNRAS, {\bf #2}, #3}
\def\pra #1 #2 #3 {#1, Phys.~Rev.~A., {\bf #2}, #3}
\def\prb #1 #2 #3 {#1, Phys.~Rev.~B., {\bf #2}, #3}
\def\prc #1 #2 #3 {#1, Phys.~Rev.~C., {\bf #2}, #3}
\def\prd #1 #2 #3 {#1, Phys.~Rev.~D., {\bf #2}, #3}
\def\pre #1 #2 #3 {#1, Phys.~Rev.~E., {\bf #2}, #3}
\def\prl #1 #2 #3 {#1, Phys.~Rev.~Lett., {\bf #2}, #3}
\def\plb #1 #2 #3 {#1, Phys.~Lett.~B., {\bf #2}, #3}
\def\science #1 #2 #3 {#1, Science., {\bf #2}, #3}
\def\nature #1 #2 #3 {#1, Nature., {\bf #2}, #3}
\def\nphysa #1 #2 #3 {#1, Nucl.~Phys.~A., {\bf #2}, #3}
\def\nphysb #1 #2 #3 {#1, Nucl.~Phys.~B., {\bf #2}, #3}
\def\nphysbs #1 #2 #3 {#1, Nucl.~Phys.~B.~Suppl., {\bf #2}, #3}

\def\h#1{\hbox{${}^{#1}$H}}
\def\he#1{\hbox{${}^{#1}$He}}
\def\li#1{\hbox{${}^{#1}$Li}}
\def\be#1{\hbox{${}^{#1}$Be}}

\def\h502{\hbox{$ h^{2}_{50}$}}

\def\fun#1#2{\lower3.6pt\vbox{\baselineskip0pt\lineskip.9pt
  \ialign{$\mathsurround=0pt#1\hfil##\hfil$\crcr#2\crcr\sim\crcr}}}
\begin{document}
\draft
\title{Observational Constraints on Dark Radiation in Brane Cosmology}
\author{ K. Ichiki$^{1,2}$, M. Yahiro$^{3}$, T. Kajino$^{1}$, M. Orito$^{1}$,
and G. J. Mathews$^{4}$ }
\address{
$^1$National Astronomical Observatory, 2-21-1, Osawa, Mitaka, Tokyo
181-8588,
Japan}
\address{
$^2$University of Tokyo, Department of Astronomy, 7-3-1
Hongo, Bunkyo-ku, Tokyo 113-0033, Japan }
\address{
$^3$Department of Physics and Earth Sciences, University of the Ryukyus,
Nishihara-chou, Okinawa 903-0213, Japan }
\address{$^4$Center for Astrophysics,
Department of Physics, University of Notre Dame, Notre Dame, IN 46556 }
\date{\today}
\begin{abstract}
 We analyze the observational constraints on brane-world cosmology
whereby the universe is described as a three-brane
embedded in a five-dimensional anti-de Sitter space.
 In this brane-universe cosmology, the Friedmann
 equation is modified by the appearance of extra terms which derive
from existence of the extra dimensions.
 In the present work we concentrate on the
 ``dark radiation'' term which diminishes with cosmic scale factor as $a^{-4}$.
 We show that, although the observational constraints from
 primordial abundances allow only a small contribution when this  term is
 positive, a much wider range of negative values is allowed.
Furthermore, such a negative contribution can reconcile the tension
 between the observed primordial $\he4$ and D abundances.
 We also discuss the possible constraints on this term
from the power spectrum of CMB anisotropies in the limit
of negligible cosmological perturbation on the brane world.
We show that BBN limits the possible contribution from
dark radiation just before the $e^+e^-$ 
annihilation epoch to lie between $-123\%$ and $+11\%$ of the 
background photon energy density.
Combining this with the CMB constraint reduces
this range to between $-41\%$ and $+10.5\%$ at the $2\sigma$ confidence level.
 \end{abstract}

\maketitle
%
%
%

\section{INTRODUCTION}

Brane-world cosmology is of considerable current interest.
In such scenarios, our universe is a submanifold embedded
in a higher-dimensional spacetime.
Physical matter fields are confined to this submanifold,
while gravity can reside in the higher-dimensional spacetime.
This  paradigm was first proposed \cite{Arkani,Shiu}
 as a means to reconcile the mismatch 
between of the scales of particle physics and gravity.
It lowers the scale of gravity to the weak scale
by introducing {\it large} extra dimensions. Although
this eliminates
the hierarchy between the weak scale and the Planck scale $M_{\rm pl}$,
it generates a new hierarchy between the weak scale and the size of the
extra dimensions. Randall and Sundrum \cite{Randall} 
have shown that a possible solution
to this new  hierarchy problem can be achieved by 
introducing noncompact extra dimensions.
This compactification is an alternative to the standard
Kaluza-Klein (KK) scheme. In their model,
our universe is described as a three-brane 
embedded  in a five-dimensional anti-de Sitter space $AdS_{5}$
(the bulk).  This guarantees the usual 4-dimensional Newtonian limit in our
brane world. 

The cosmological evolution of such brane universes has been extensively
investigated.  Exact solutions have been found
by several authors \cite{Kraus,Binetruy,Hebecker,Ida,Mukohyama,Vollick}.
These solutions reduce to a generalized Friedmann equation on our brane
which can be written  as
\begin{equation}
\left(\frac{\dot{a}}{a}\right)^2
=\frac{8 \pi G_{\rm N}}{3} \rho
-\frac{K}{a^2}+\frac{\Lambda_{4}}{3}
+\frac{\kappa_{5}^4}{36}\rho^2 + \frac{\mu}{a^4}~~.
\label{Friedmann}
\end{equation}
Here,  $a(t)$ is the scale factor at cosmic time $t$, and
$\rho$ is the total energy density of matter on our brane.

In equation (\ref{Friedmann}), several identifications of cosmological parameters 
were required in order to recover standard big-bang cosmology.
For one, the first term on the right hand side
is obtained by relating 
the four-dimensional gravitational constant $G_{\rm N}$
to the five-dimensional gravitational constant, $\kappa_{5}$.  Specifically,
\begin{equation}
G_{\rm N} = \kappa_{5}^4 \lambda / 48 \pi~~,
\end{equation}
 where $\lambda$ is the intrinsic tension of the brane
and $\kappa_5^2 = M_5^{-3}$, where $M_5$ is the five dimensional Planck mass.
Secondly, the four-dimensional cosmological constant $\Lambda_{4}$
is related to its five-dimensional counterpart $\Lambda_{5}$,
\begin{equation}
\Lambda_{4}=\kappa_{5}^4 \lambda^2 /12 + 3 \Lambda_{5}/4~~.
\end{equation}
$\Lambda_{5}$ should be negative in order for $\Lambda_{4}$ to obtain
its presently observed small value.

Standard big-bang cosmology does not contain the fourth and fifth terms
of Eq. (\ref{Friedmann}).
The fourth term arises from the imposition of a  junction condition
for the scale factor on the surface of the brane. Physically, it 
derives from a singular behavior in the energy-momentum tensor
which originates in the fact that physical matter fields are confined to the brane.
This $\rho^2$ term would decay rapidly as $a^{-8}$ in the early radiation
dominated universe.  Hence,  it is not likely to be significant
during the later nucleosynthesis and photon decoupling epochs of interest here.

The fifth term, however, is of considerable interest for the present
discussion. It scales just like radiation with a constant $\mu$.
Hence, it is called the dark radiation. 
This term derives from the electric (Coulomb) part of
the five-dimensional Weyl tensor \cite{Shiromizu}.
The coefficient $\mu$ is a constant of integration obtained by
integrating the five-dimensional Einstein equations 
\cite{Kraus,Binetruy,Hebecker,Ida,Mukohyama,Vollick}.
Both positive and negative $\mu$ are possible mathematically.
Its magnitude and sign  can depend on the choice of initial conditions when
solving the five-dimensional Einstein equation.
Hence, even the sign of $\mu$ remains an open question \cite{Maeda}.

Dark radiation should strongly affect both 
big-bang nucleosynthesis (BBN)
and the cosmic microwave background (CMB).
Hence, such observations can be used to constrain both the 
magnitude and sign of the dark radiation.
A brief analysis of this  was made in literature \cite{Binetruy}.
In the present work, we seek to explore these constraints 
in more detail utilizing the most recent light-element 
abundance constraints 
and the latest combined CMB power spectrum data sets.

\section{BBN constraint}
The observed primordial light-element abundances 
constrain the conditions during the BBN epoch from the time
of weak reaction freezeout ($t \sim 1$ sec, $T \sim 1$ MeV) to the
freezeout of nuclear reactions  ($t \sim 10^4$ sec, $T \sim 10 $ keV).
The present status the observational constraints have been reviewed 
in a number of papers (cf.~\cite{Olive}-\cite{pinns}).
The primordial helium abundance  is obtained by measuring 
extragalactic $\rm H{II}$ regions in low-metallicity irregular 
galaxies.  The primordial heluim abundance $Y_p$ so deduced
 tends to reside in one of two possible values (a low value
$Y_p \approx 0.230$,  \cite{Olive} and a high value $Y_p \approx 0.245$
\cite{Thuan}).
In view of this controversy, 
we adopt a conservative range for the primordial  $\he4$
abundance:
\begin{equation}
0.226\leq Y_p \leq 0247.
\end{equation}

Primordial deuterium is best determined from its absorption line
in  high redshift Lyman $\alpha$ clouds 
along the line of sight to background quasars.
For deuterium there is a similar possibility for either a high or
low value.  For the present discussion,
however, we shall adopt the generally
accepted low value for D/H \cite{tytler00,omera}.
\begin{equation}
2.9\times 10^{-5}\leq {\rm D/H} \leq 4.0\times 10^{-5} \hspace{0.5cm}
\end{equation}

The primordial lithium abundance is generally
inferred from old low-metallicity halo stars.
Such stars exhibit an approximately constant (``Spite plateau'') 
lithium abundance as a function of surface temperature
which is taken to be the primordial abundance. 
There is, however, some controversy \cite{pinns} concerning
the depletion of $\li7$ on the surface of such halo stars. 
If destruction has occurred, the true primordial $\li7$
abundance is higher than the plateau value.  For the
present purposes, therefore, we adopt a
conservative $\li7$ abundance constraint:
\begin{equation}
1.67\times 10^{-10} \leq \li7{\rm/H} \leq 4.75\times 10^{-10}.
\end{equation}

The constraints on positive extra energy density during the BBN epoch
based upon primordial light-element abundances
have been recently studied by many authors in
context of numbers of neutrino families,
lepton asymmetry, or  dark energy (cf.~\cite{Olive,Orito,Yahiro}). 
The main effect of such additional background 
energy density is to increase the
universal expansion rate. This causes the neutron to proton ratio
to be larger because the weak reactions freeze out at a higher temperature and
because there is less time for neutrons to decay 
between the time of weak-reaction
freezeout and the onset of BBN.
Consequently, adding excess energy density during 
BBN yields a larger $\he4$ abundance since most of the free
neutrons are converted into $\he4$ nuclei. 
D/H also increases largely because the reactions destroying deuterium fall
out of nuclear statistical equilibrium while the deuterium abundance is 
higher \cite{Smith}.
Similarly, there is less time for the destructive reaction
$\li7(p,\alpha)\he4$.  This causes $\li7$ to be more abundant for
$\eta \leq 3\times 10^{-10}$.  However,
there is also less time for the  $\he4 (\he3,\gamma)\be7$ reaction to occur.
This causes $\li7$ to be less abundant for
$\eta \geq 3\times 10^{-10}$. 
On the other hand, when the extra energy
component is negative (i.e. negative dark radiation), the opposite
results occur.

\begin{figure}
\rotatebox{-0}{\includegraphics[width=0.45\textwidth]{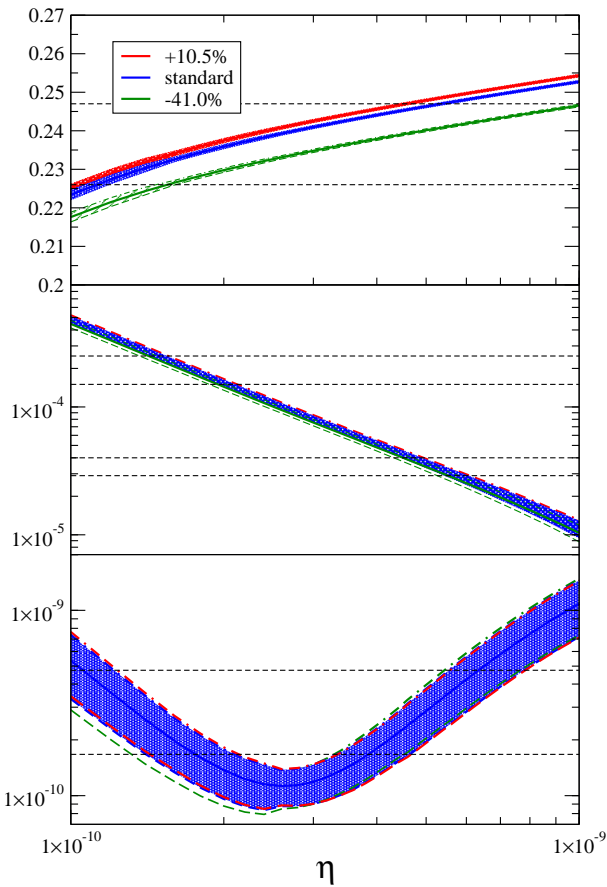}}
\caption{Light-element abundances  as a function of baryon to photon ratio $\eta$.
 Shaded Areas or dashed lines  denote $\pm 2\sigma$ uncertainties in the BBN model predictions. 
Plotted are models with 0\% (blue), +10.5\% (red) and -41\%
 (green) dark radiation
 (relative to the background photon energy density just before the
$e^+e^-$ annihilation epoch).  
$\he4$, D and $\li7$ are shown in the top, center and bottom panels,
 respectively. The \he4 abundance predictions are well separated for the
three dark radiation models, while the models are barely distinguishable
for D and \li7.  Observational constraints are indicated as horizontal lines as labeled.
}
\label{fig:1}
\end{figure}

Figure \ref{fig:1} illustrates the dependence of the nucleosynthesis yields
with the dark radiation content. In the following, we will quote the
dark radiation content as a fraction of the background photon energy
density just before and the onset of the $e^+e^-$ annihilation and BBN epochs.
We have included the current 2$\sigma$ uncertainties \cite{nollett00} arising
from the input nuclear reaction rates in our present analysis of the BBN
model predictions.

For dark radiation in the range of 0 to $+11\%$ of the background 
photon energy density,
the cosmological bounds on the baryon-to-photon ratio  $\eta$
come from the $\he4$ upper bound and the D/H upper bound.
With negative dark radiation the allowed range for $\eta$
expands because the $\he4$ mass fraction and D/H
have opposite dependences on $\eta$.
The addition of more than 2\% negative dark radiation reduces the expansion rate
and the helium abundance sufficiently so that the adopted $\he4$
constraint is satisfied for all values of $\eta$ 
which satisfy the D/H constraint.
For negative dark radiation 
in the range of $2\%-112\%$
of the background 
photon energy density, the constraint  on $\eta$ comes only
from D/H upper and lower limits.  
Between 112 and 123\% negative dark radiation,
the constraint on $\eta$ comes from the lower bounds on $\he4$ 
and D/H.

Similarly, the conservative  $\li7$ abundance constraint adopted here
does not significantly constrain the dark radiation component.
The shaded region on Figure \ref{fig:2} shows 
 allowed values of the dark radiation fraction,
$\rho_{DR}/\rho_\gamma$, where $\rho_\gamma$  is the total energy
density in background photons
just before the BBN epoch at $T = 1$MeV.  Note, that
only a small ($\le 11\%$) positive dark radiation contribution is allowed
while substantial negative dark radiation (up to 123\%) is allowed
and even preferred by the BBN constraints.

\section{CMB Constraint}
Next we examine the possible imprint of  a dark radiation on the 
CMB angular power spectrum.
It is well known that the CMB spectrum is sensitive to many 
cosmological parameters which have almost no effect on BBN.
For simplicity, therefore, we
have fixed most cosmological parameters to their optimum
values and explore the effects of varying the 
dark radiation content and baryon to photon ratio $\eta$.
In spite of the name dark ``radiation'' it has no interaction 
(e.g.~Compton scattering) with other
matter fluids.
Moreover,  we make the further simplifying assumption
that it has no intrinsic fluctuation. 
Cosmological perturbation theory in a five dimensional universe is now
extensively under consideration \cite{Sasaki}-\cite{Koyama}.
Ultimately, one must take the five dimensional (geometrical) perturbative
effects into account when calculating the CMB angular power
spectrum.  In this present paper, however, we only address the dominant effect
on the background expansion rate.

\begin{figure}
\rotatebox{-90}{\includegraphics[width=0.33\textwidth]{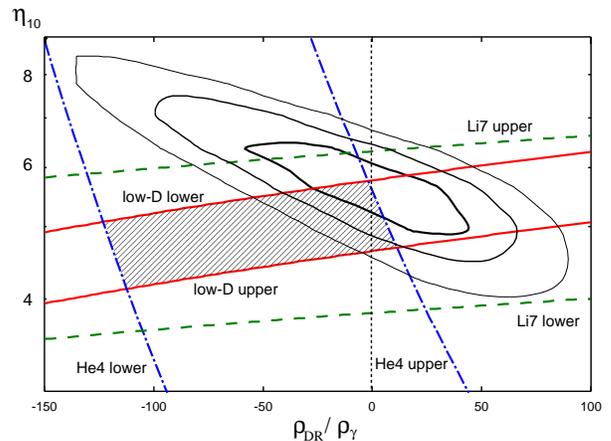}}
\caption{Constrains from the primordial abundances (lines)
and the CMB (contours) 
on  $\eta$ and the fraction of the background photon energy density
in dark radiation $\rho_{DR}/\rho_\gamma$
just before the $e^+e^-$ annihilation.  
Lines are labeled by the light-element constraints corresponding
 to the central values in Fig. 1.  The shaded region denotes the  parameters 
allowed by BBN.  The contours show the 1, 2, and $3\sigma$ limits
from fits to the CMB power spectrum.
 }
\label{fig:2}
\end{figure}

We have calculated 
CMB power spectra 
using the CMBFAST code of \cite{Seljak}. 
The $\chi^2$ goodness of fit  
to the combined BOOMERANG \cite{boom}, DASI \cite{dasi}, 
and MAXIMA-1 \cite{max} data sets was evaluated using 
the widely employed offset log-normal procedure of  
\cite{Bond}.  The available experimental offsets and window functions
were utilized.  As a benchmark zero dark-radiation
model, we  have made a search for a global optimal
fit to the combined CMB data set
for a flat $\Omega_M + \Omega_\Lambda = 1$ cosmology
(with ionization parameter $\tau = 0$).  
We find a best fit for
$\Omega_M = 0.233$, $\Omega_\Lambda = 0.767$,
$h = 0.726$, $\Omega_b h^2 = 0.0214$ ($\eta_{10} = 5.75$), $n = 0.9334$.
We have also marginalized over the experimental calibration
uncertainties and the {\it COBE} normalization using Gaussian priors
based upon published experimental uncertainties.
This fit  gives a $\chi^2$ of 30.03 for 31 degrees of freedom.
For the present purposes, we restrict our consideration to
this parameter set
as an optimum 4-dimensional standard cosmology.
We then study variations in the goodness of fit 
as a function of the dark radiation fraction
at the photon decoupling epoch and $\Omega_b h^2 = \eta_{10}/268$.
In addition, the normalization was optimized for 
each choice of these two parameters.

The most distinguishable effects of the dark radiation is
their influence on the location and amplitude of
the  acoustic peaks in the CMB power spectrum \cite{Hu1}.
Adding positive dark radiation moves the epoch of  matter radiation equality
to a later epoch. 
It prevents the growth of perturbations inside the horizon 
and leads to a decay in the gravitational potential.
This increases the amplitude of the CMB acoustic oscillations
by the integrated Sachs-Wolfe effect.
The net result of adding
positive dark radiation  is therefore an enhanced CMB anisotropy.
The opposite is true if negative dark radiation is added. 

As  a second feature, a more 
rapid expansion rate  due to positive dark radiation
causes the epoch of photon decoupling
to occur earlier so that the horizon size
is smaller at the surface of photon last
scattering.  Therefore the $l$-values for the acoustic peaks are 
slightly larger or smaller depending upon whether the dark radiation
term is positive or negative.

\begin{figure}
\rotatebox{0}{\includegraphics[width=0.45\textwidth]{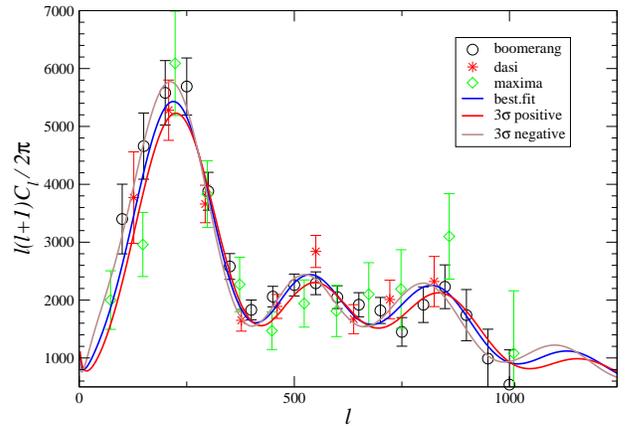}}
\caption{An optimum fit to the CMB angular power spectrum 
compared with fits obtained by adding a $3\sigma$ negative or positive dark
 radiation component.
}
\label{fig:3}
\end{figure}

Figure \ref{fig:3}
 illustrates effects of both positive and negative dark radiation on
the CMB angular power spectrum.  
This shows the bench-mark zero dark-radiation model together with 
$\pm 3\sigma$ components of dark radiation.  These limits correspond
to a ratio of dark radiation to photon energy density
of $+24\%$ and $-35\%$ at the 
photon decoupling epoch.   Correcting for photon heating at the pair annihilation
epoch, these limits expand by a factor of $(11/3)^{4/3} = 3.85$
for the dark-radiation fraction just before the BBN epoch.  
Hence, these limits would be +92\% and -135\% of the photon energy density
just before nucleosynthesis.

The final effect on the power-spectrum depends upon the
normalization, and is somewhat counter intuitive.
From Figure \ref{fig:3} we see, for example,  that 
a fit with a dark radiation fraction
of -35\% of the photon energy density at the CMB epoch 
increases (rather than decreases) the amplitude of the
first acoustic peak by $\approx$10\%  and shifts the
location of the first and third peaks
to smaller $l$ values.  The increase in the acoustic peak amplitudes
is a result of having shifted the normalization to optimize the 
goodness of fit.  Note, that the effects
of positive or negative dark radiation are not the same as
simply adding or subtracting photons.  This is because
dark radiation does not behave like relic photons or neutrinos.
Dark radiation does not interact either gravitationally or via Compton scattering
with the other matter fields.  Also, in the present analysis, it does not
fluctuate. Hence, the effects of dark radiation on
the power spectrum are 
in principle distinguishable from the effects of normal 
electromagnetic radiation.

Figure \ref{fig:2} shows the contours of the dark radiation fraction
and $\Omega_b h^2$ allowed by nucleosynthesis and the CMB.
This shows that the combined nucleosynthesis and CMB constraints 
severely limit the possible sign and amplitude of the dark radiation.
The combined $2\sigma$ 95\% confidence limit from the
concordance of both constraints
corresponds to $-41\% \le \rho_{DR}/\rho_\gamma \le +10.5\%$
for $4.73 \le \eta_{10} \le 5.56$ (or $0.0176 \le  \Omega_b h^2 \le 0.0207$).


\section{discussion}
We have considered the cosmological constraints on 
the magnitude and sign of the dark radiation term of the brane-world
generalized Friedmann equation (\ref{Friedmann}).
If the sign of the dark radiation is positive 
then it behaves like additional relativistic
particles and enhances the expansion rate. 
This kind of effect has been recently well studied \cite{Yahiro} in 
the context, for example,  of additional neutrino flavors or 
degeneracy and is tightly constrained.
We have reexamined this effect  for both positive and negative
dark energy.  We include the nuclear reaction
uncertainties in the BBN model predictions.  For positive dark radiation
the observational upper bound for primordial $\he4$ and D/H allows at
most $\rho_{DR}/\rho_{B} \leq 0.03$ ($\rho_{DR}/\rho_\gamma \leq 0.16$)
at the BBN epoch.  This limit is consistent with the estimate 
(
 on thermally generated dark radiation due to bulk graviton production
described in \cite{Hebecker}.

Such extra energy also affects the power spectrum 
of  CMB fluctuations, but it is too
small to be constrained by current CMB measurements.  We therefore conclude
that BBN places the most stringent constraint on positive dark
radiation.

On the other hand, a wider
range of dark radiation density relative to the background photon energy density
is allowed in the case of negative dark radiation. 
We deduce an absolute BBN upper limit of 123\% negative dark
radiation. This maximal value is allowed for $\eta \approx 5.09\times
10^{-10}$  (or $\Omega_b h^2 \approx 0.019$). This $\eta$
value, however, is $1\sigma$ less than the values consistent
with the combined BOOMERANG, DASI, and MAXIMA-1 data sets. 
For the combined CMB and BBN analysis, we deduce that only a
maximum of 41\%
contribution of negative dark radiation is allowed at the 95\% 
confidence level.

We should, however, point out several caveats to the present work.
One is  
that if one wishes to avoid a naked singularity in the
bulk dimension, then there is a relation
\cite{Maeda} between the curvature $K$ and
the dark radiation $\mu$ when the sign of $\mu$ is negative,
\begin{equation}
\mu \geq -\frac{K^2 l^2}{4}\hspace{1cm} ~,~\mbox{for } \mu<0~~,
\label{relation}
\end{equation}
where $l$ is the five dimensional curvature length scale which
relates to the five dimensional cosmological
constant $\Lambda_5 \equiv -{4 / l^2}$.
If one accepts this cosmic censureship hypothesis, then only minuscule
quantities of negative dark radiation are allowed if one wishes
to maintain the five dimensional Planck mass above the TeV scale.
Hence, the present limits on negative dark radiation only apply
if one wishes to accept either  a much lower value for $M_5$ or a naked
singularity in the bulk dimension.

We note that there is an independent constraint 
\cite{cline} on the
five-dimensional Planck mass from the quadratic term in equation
(\ref{Friedmann}).  If one applies the same BBN constraint that not
more than 3\% of the background energy density can be in this
term at the time of weak-reaction freezeout, then the condition
\begin{equation}
{(\kappa_5^4 \rho^2/36)  \over (8 \pi G_N/3)} \le 0.03 \rho~~,
\end{equation}
implies a limit of $M_5 \ge 10$ TeV.

As another caveat we note that ultimately this analysis should be
repeated with the inclusion of cosmological perturbations
in the dark radiation.
The evolution of cosmological perturbations in a brane universe is now being
investigated by many authors \cite{Sasaki}-\cite{Koyama}.
The main difficulty with this subject comes from the fact that even four
dimensional perturbation equations on our brane have fifth
dimensional curvature corrections. Ultimately, one must solve the
cosmological perturbations in the bulk.
Langlois \cite{Langlois} has proposed that these perturbations from the bulk
appear like source terms in the four dimensional perturbation equations.
They are similar to ``active seeds'' in the context of topological
defects.  Therefore, it seems unlikely that these source terms could
provide  a dominant contribution to the CMB anisotropies.
In this discussion, therefore,  we have ignored these perturbations and
concentrated on the zeroth order effect of ``dark radiation'' which
appears in equation (1).

Finally we note that although 
the the acoustic peaks are very useful indicators of 
the dark radiation, they are also sensitive to other
cosmological parameters, especially $\Omega_m$ and $\Omega_\Lambda$.
Hence, one should ultimately do a combined likelihood
analysis including other constraints on
cosmological parameters to test for the
significance of the dark radiation component in the CMB.

\section{Conclusion}
We conclude that the constraints on BBN alone allow for
$-1.23 \leq \rho_{\rm DR}/\rho_\gamma \leq 0.11$ in 
a dark radiation component. 
In order to compute the theoretical prediction of CMB anisotropies
exactly, one must eventually solve the perturbations including the contribution
from the bulk.  
However, we have shown that the CMB power spectrum 
can be used to constrain the dominant expansion-rate effect of 
the dark-radiation term in the generalized Friedmann equation.
If the constraint from effects of dark radiation on the expansion rate
are included, the allowed concordance range
range of dark radiation content reduces to
$-0.41 \leq \rho_{\rm DR}/\rho_\gamma \leq 0.105$ at the 95\% confidence level.
\acknowledgements
One of the authors (MY) would like to thank K. Ghoroku for helpful discussions on
the theoretical implications of dark radiation.
This work has been supported in part by the Grant-in-Aid for
Scientific Research (10640236, 10044103, 11127220, 12047233)
of the Ministry of Education, Science, Sports, and Culture of Japan.
Work at the University of Notre Dame was supported
by the U.S. Department of Energy under 
Nuclear Theory Grant DE-FG02-95-ER40934.

\end{document}